# Wavelets for Single Carrier Communications

Ömer Bulakci

PhD candidate at Aalto University School of Electrical Engineering, *omer.bulakci@ieee.org*

**Abstract**: This paper aims to provide a report regarding the seminar presentation given on 23.02.2011 as a part of the postgraduate seminar course *S-88.4223 Wavelets in Communications* [1] lectured by Dr. Sumesh Parameswaran at Aalto University School of Electrical Engineering. In particular, the topic on "wavelets for single carrier communications" has been considered in this paper.

## I. Introduction

Wavelets are small oscillatory waveforms that are non-zero for a limited period of time, satisfying certain mathematical conditions. The primary advantages of wavelet transform are considered to be *wavelet diversity* and *high localization in time and frequency domains*. Fast wavelet transform is implemented via filter banks. One important feature of the wavelets is that wavelet symbols are overlapping both in time and frequency. This feature enables a high spectral efficiency. Another important feature of the wavelets is the so-called *wavelet diversity*. Wavelet diversity implies that there exist various wavelet families and new wavelets can be designed depending on the application of interest. This paper provides an overview of the wavelet applications in the field of single carrier communications.

## II. Wavelet Modulation

It is promising to use bandwidth efficient overlapped wavelet packet functions for modulation. It has been shown that the orthogonality of the wavelets can be exploited to increase the bandwidth efficiency [1]. Since wavelet packet functions are overlapped a guard band consisting of cyclic prefix is not available for the wavelets with a multi-channel scheme. The advantage of this is the increased bandwidth efficiency at the cost of decreased flexibility against multipath fading tolerance due to the lack of a cyclic prefix.

In [2], two wavelet systems are proposed. In the first system, wavelets are used for pulse shaping. It has been shown that wavelet based pulse shaping (with Daubechies 10) the spectral efficiency of the system can be made higher than the system using raised cosine. It is further shown that the spectral efficiency can be increased by means of using mother wavelet together with the dyadic expansions. Note that in [3], the author shows that the upper limit of the spectral efficiency is two, if infinite dyadics are used. For the first system it is also demonstrated that different wavelet families can have different spectral efficiencies (see Table 2). The spectral efficiencies are tabulated in Table 1 and Table 2 [2]. In Table 1, wavelet 1, wavelet 1.5, and wavelet 1.75 correspond to mother wavelet, mother wavelet plus first dyadic expansion and mother wavelet plus first and second dyadic expansions, respectively. In the second proposed system, the classical oscillator is replaced by a waveform generator that produces a basic mother wavelet [2]. This model is referred to as Wavelet Shift Keying (WSK). The authors show that the spectral efficiency of this system becomes comparable (even better) with the BPSK system using raised cosine filtering when other user's information is as well transmitted on the same band. Besides, the authors mention that the binary WSK performs similar to BPSK in the presence of AWGN.

---

[1] https://sites.google.com/site/tkksigpostgrad/Home/wavelets-in-communication-engineering

**Table 1. Comparison of SE of RC with Daubechies system**

| System | Spectral Efficiency (Base band) b/S/Hz | Spectral Efficiency (Pass band) b/S/Hz |
|---|---|---|
| RC | 0.83 | Not Applicable |
| Wavelet 1 | 0.7 | 0.85 |
| Wavelet 1.5 | 1.01 | 1.10 |
| Wavelet 1.75 | 1.12 | 1.22 |

**Table 2. Comparison among different Wavelet Systems**

| Wavelet System (Using three dyadics) | Spectral Efficiency (Base band) b/S/Hz | Spectral Efficiency (Pass band) b/S/Hz |
|---|---|---|
| Daubechies | 1.12 | 1.22 |
| Symlets | 1.05 | 1.14 |
| Coiflets | 0.95 | 1.04 |

### III. Bandwidth Efficiency (Demonstration)

In the presentation, one of the strong aspects of the wavelets has been presented. Namely, the wavelet diversity can be exploited to find a wavelet function with desired bandwidth characteristics. By means of the spectra of different QMFs (as used in the filter banks) corresponding to different wavelet families, it is shown that the spectra of these wavelet families can differ significantly. Furthermore, the impact of the receiver bandwidth on the correct detection of the receiver symbols has been demonstrated. In this demonstration, the error vector magnitude (EVM) between the received symbol and ideal symbol is determined with respect to the normalized bandwidth of the receiver filter. In this scheme, Fourier transform based Orthogonal Frequency Division Multiplexing (OFDM) and wavelet transform based OFDM are as well compared using 512 subcarriers and an oversampling factor of two. For the wavelet transform based OFDM Daubechies 10 (db10) wavelet is used. It is observed that wavelet transform based OFDM is more tolerant to the inband distortion, which is supported by the fast decay of the side-lobes. It is worth noting that through different wavelet families such performance can be further enhanced. The trade-off is in general between the complexity of the wavelet implementation and the performance.

### IV. The modified Gaussian: A Novel Wavelet

One of the key points of the wavelet based communications is the inherent flexibility of choosing the basis functions. Moreover, different wavelets can be as well designed. In [4] a novel wavelet called *the modified Gaussian* is introduced. It is presented that the new wavelet has very low side lobes and its bandwidth efficiency is compared to that of the square-root raised cosine. This novel wavelet is obtained by applying the so-called orthogonalization trick. A further advantage of this design is that the spectral efficiency can be controlled by a parameter. The Fourier transform of the scaling function is formulated in (1).

$$\Phi(f) = \frac{e^{-\sigma^2 T^2 (2\pi f)^2}}{\sqrt{\sum_{l \in \mathcal{Z}} e^{-8\sigma^2 T^2 \pi^2 (f+l/T)^2}}} \qquad (1)$$

## V. Wavelets in Single Carrier (SC)-FDMA Sytems

OFDM is a modulation technique well known from wireless and wired communications and part of standards as LTE, WiMAX, WLAN, DVB and DAB [5][6]. However, a major drawback of OFDM is high peak-to-average power ratio (PAPR). Therefore, a high dynamic range of power amplifiers and DA/AD-converters is required for distortionless transmission. The effect of high PAPR becomes deleterious for higher number of subcarriers (SCs), since PAPR of an OFDM symbol can be up to *10·log$_{10}$(N)* in the worst case, given *N* as the number of subcarriers. Being a multicarrier modulation scheme wavelet transform based OFDM (WT-OFDM) can also suffer from the high PAPR depending on the used wavelet family. In [7], the PAPR performance of the WT-OFDM is compared with that of the Fourier transform based counterpart. It has been shown that WT-OFDM has better PAPR performance and its performance increases as the wavelet order increases. Note that therein the Daubechies wavelet family has been analyzed. For the OFDM signal generation 128 SCs were used with QAM modulated input symbols.

In LTE, a DFT precoding is used before generating the FT-OFDM symbol. It is worth noting that precoding is a PAPR reduction method where the side lobes of the auto-correlation function of the input symbols are tried to be reduced. The upper limit of the PAPR is related to these side lobes. The cumulative effect of using DFT precoding and FT-OFDM is the single carrier-like modulation which is referred to as SC-FDMA [8].The SC-FDMA has superior PAPR performance relative to OFDMA and hence is preferred for the uplink (UL) transmission. Low PAPR implies high power efficiency which is critical for the mobile phones especially when we consider today's smartphones. It has been shown that similar to FT-OFDM based SC-FDMA, WT-OFDM based SC-FDMA can be designed [9]. The idea is applying DWT in advance to WT-OFDM symbol generation. The PAPR performances of FT-OFDM and WT-OFDM based SC-FDMA are compared using Haar wavelet with 9-levels (or 512 SCs for FT-OFDM) with QPSK modulated symbols. It is shown that the PAPR performances of conventional FT-OFDM and WT-OFDM are similar. On the other hand, the PAPR performances can be significantly enhanced when SC-FDMA is used. Moreover, the PAPR performance of the WT-OFDM based SC-FDMA is better than that of the FT-OFDM based SC-FDMA. In addition, the BER performance of WT-OFDM based SC-FDMA is observed to be better for a 10-path fading environment. Therefore, wavelets are also promising for SC-FDMA systems.

# Wavelets for Single Carrier Communications

**Ömer Bulakci**

\* Nokia Siemens Networks
\*\* Aalto University School of Electrical Engineering

**23.02.2011**

Seminar on *Wavelets in Communications*

# Content

- Introduction

- Wavelet Modulation

- Bandwidth Efficiency
    - Spectra of Different Wavelet Families
    - Impact of Receiver Filter Bandwidth

- The Modified Gaussian: A Novel Wavelet

- Wavelets in SC-FDMA

- Conclusion

# Introduction (1/3)

## *Wavelet Principles*

- A wavelet ψ(t) is a time limited oscillatory waveform satisfying certain mathematical conditions
- Wavelet transform performs signal analysis by means of scaled and translated versions of the mother wavelet ψ(t)

**Basis Functions**

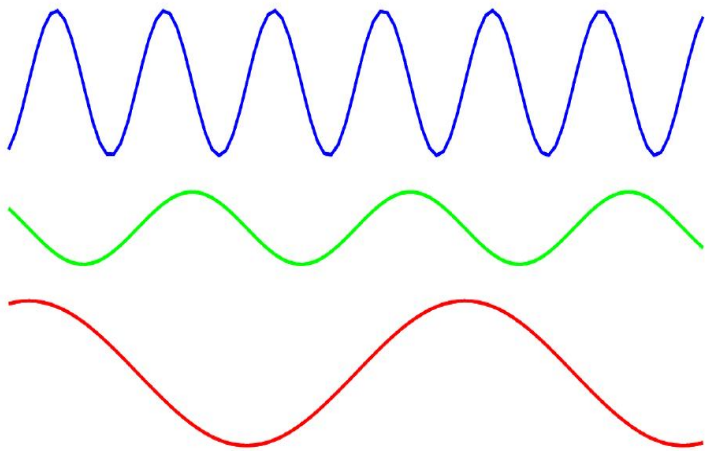

**Sine Waves**

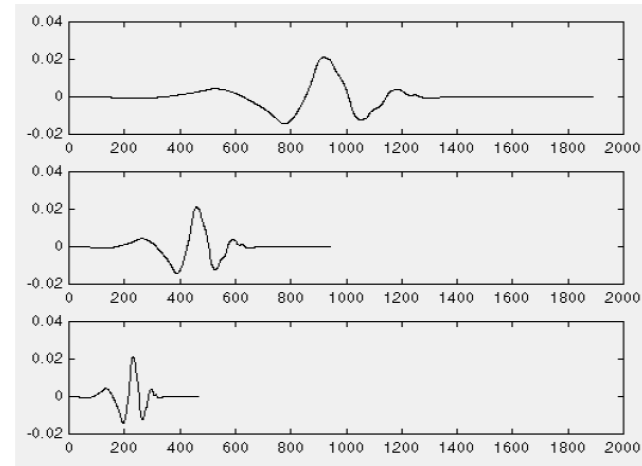

$f(t) = \psi(t)$

$f(t) = \psi(2t)$

$f(t) = \psi(4t)$

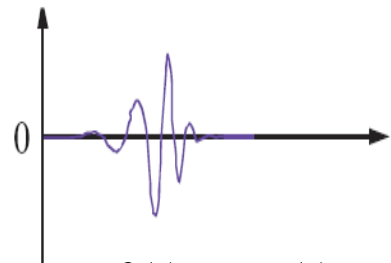

$f(t) = \psi(t)$

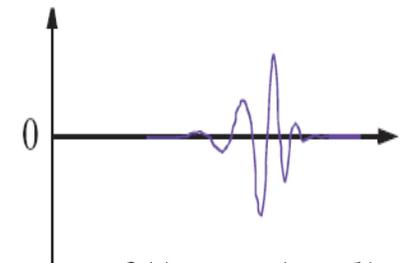

$f(t) = \psi(t-k)$

# Introduction (2/3)
## *Wavelet Principles*

- A wavelet $\psi(t)$ is a time limited oscillatory waveform satisfying certain mathematical conditions
- Wavelet transform performs signal analysis by means of scaled and translated versions of the mother wavelet $\psi(t)$
- Fast Wavelet Transform via Filter Bank implementation

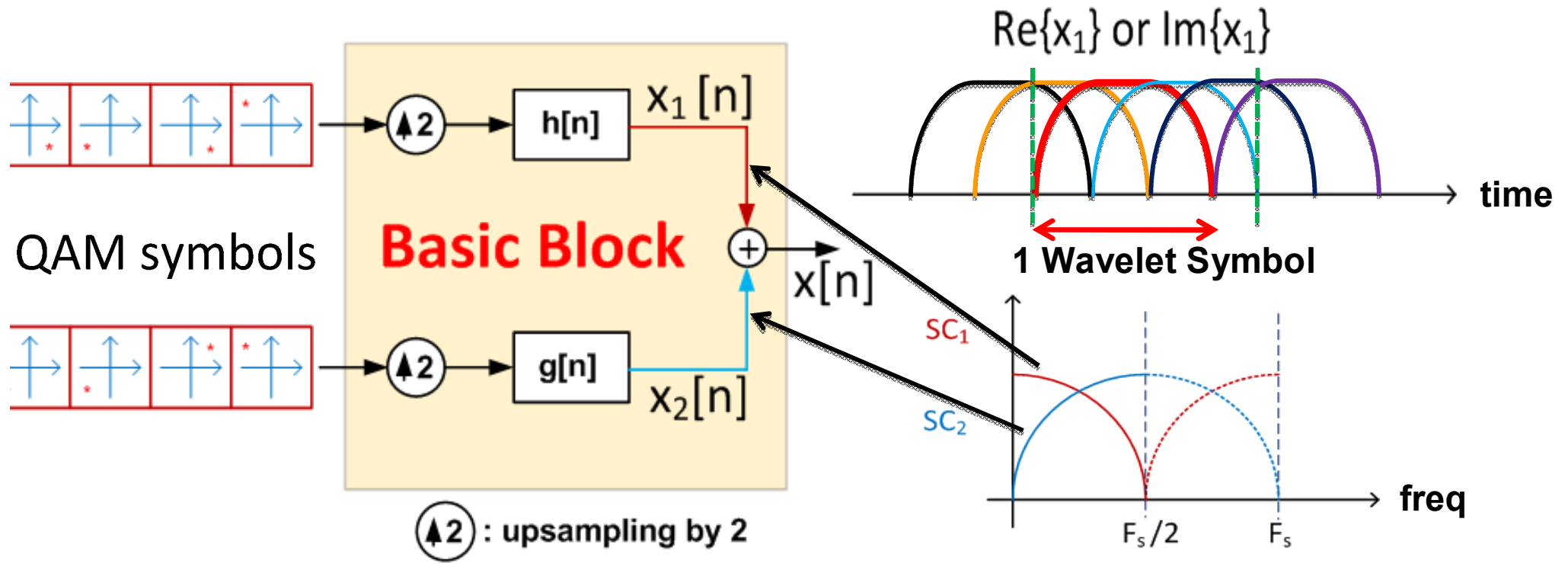

# Introduction (3/3)

## *Wavelet Principles*

- A wavelet $\psi(t)$ is a time limited oscillatory waveform satisfying certain mathematical conditions
- Wavelet transform performs signal analysis by means of scaled and translated versions of the mother wavelet $\psi(t)$
- Fast Wavelet Transform via Filter Bank implementation
- Different wavelet families exist } wavelet diversity
- New wavelets can be designed
- Convention:

wavelet family — **Johnston** **64** . **5** — design group

wavelet order

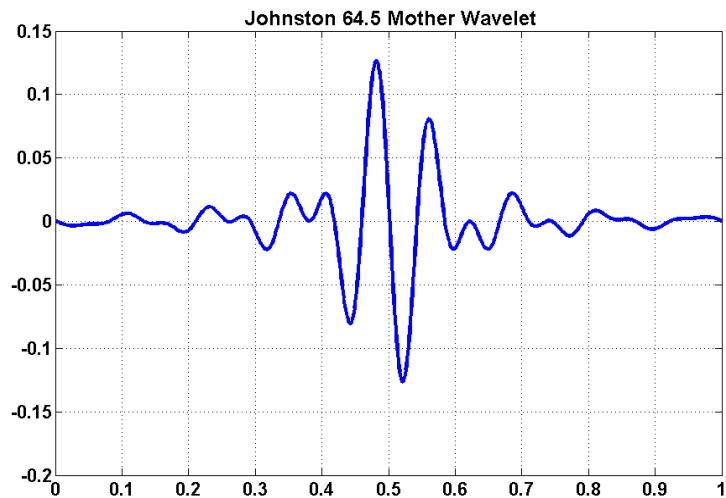
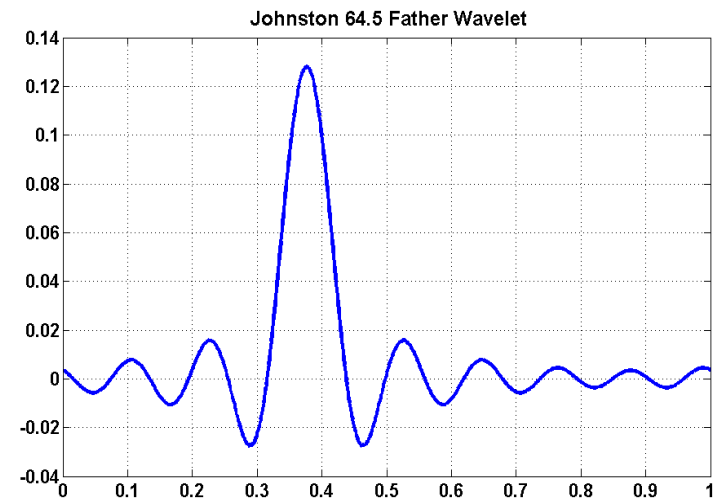

- Introduction

- **Wavelet Modulation**

- Bandwidth Efficiency
    - Spectra of Different Wavelet Families
    - Impact of Receiver Filter Bandwidth

- The Modified Gaussian: A Novel Wavelet

- Wavelets in SC-FDMA

- Conclusion

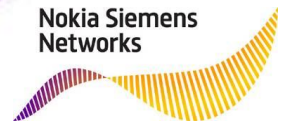

# Wavelet Modulation (1/3)

- Bandwidth efficient overlapped wavelet packet functions can be used for modulation.
- As the wavelet packet functions are orthogonal to each other a *guard band* is not used.

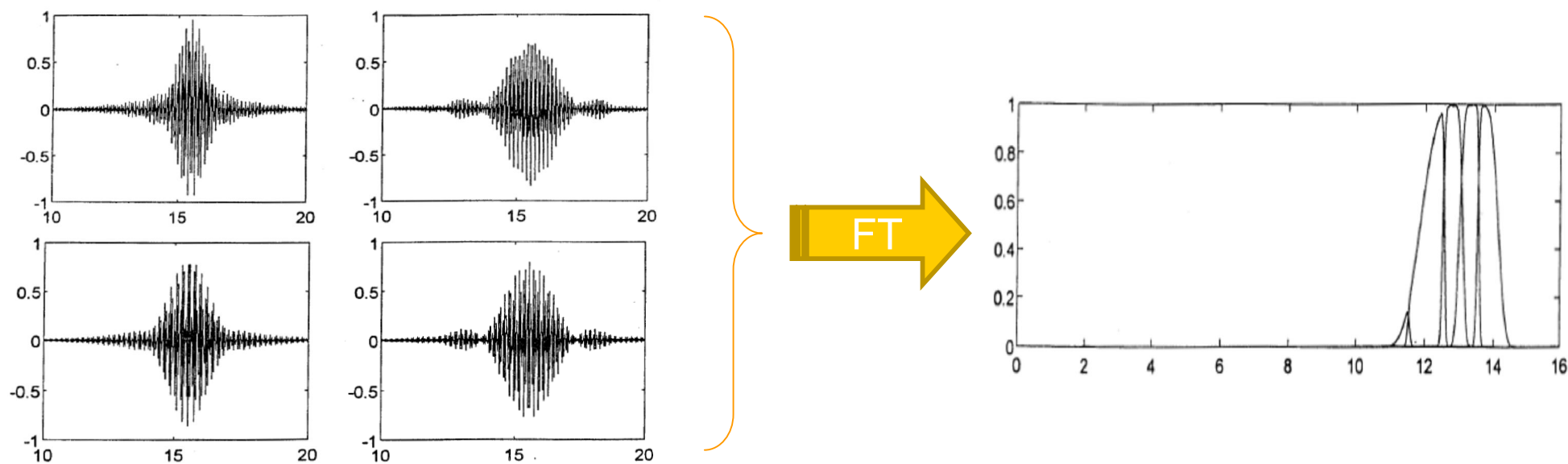

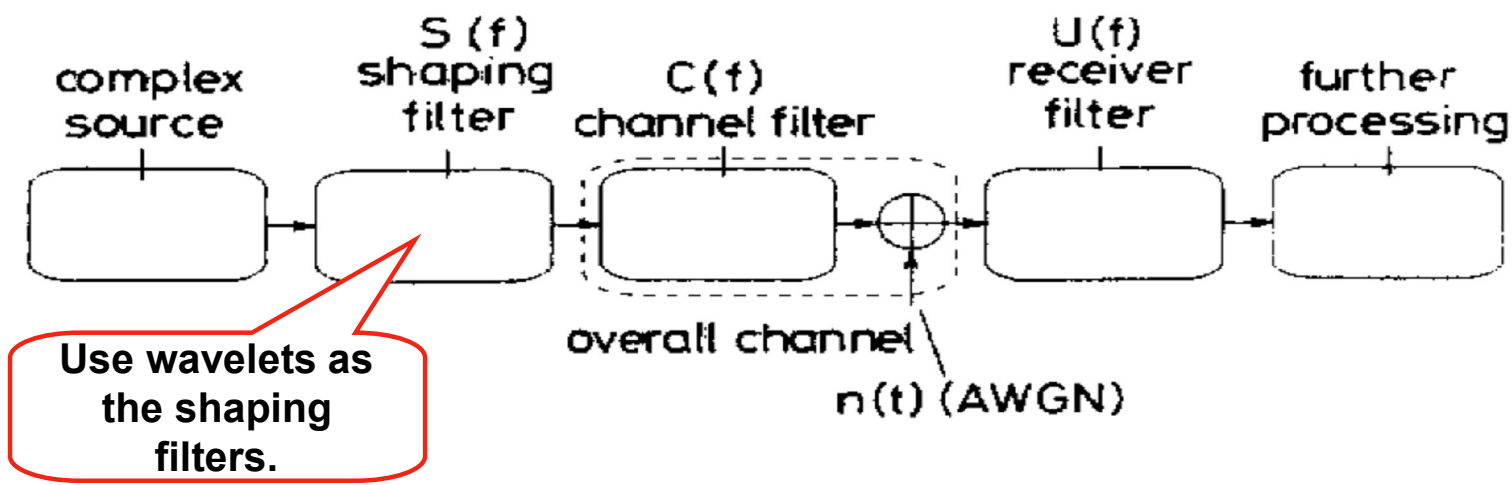

**Use wavelets as the shaping filters.**

# Wavelet Modulation (2/3)
## System 1: Pulse Shaping

- Wavelets and their dyadic translates are used as baseband shaping pulses to increase the spectral efficiency.
- Higher spectral efficiency is attained compared to raised-cosine (RC) signaling.

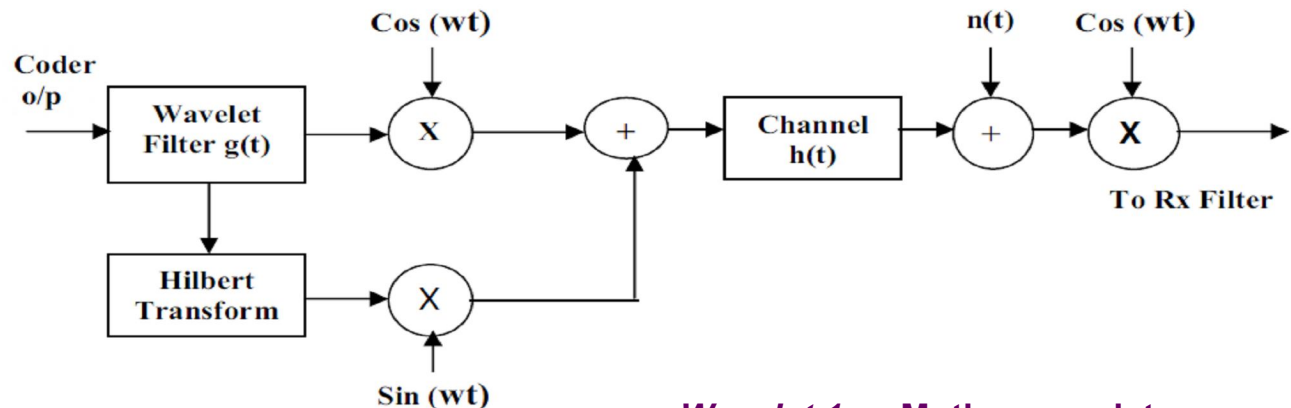

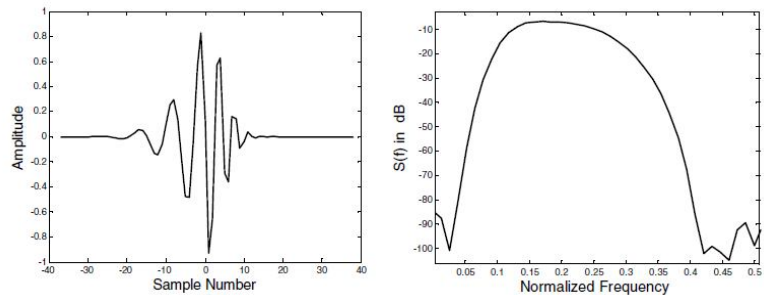

(a) Mother wavelet (b) Its PSD

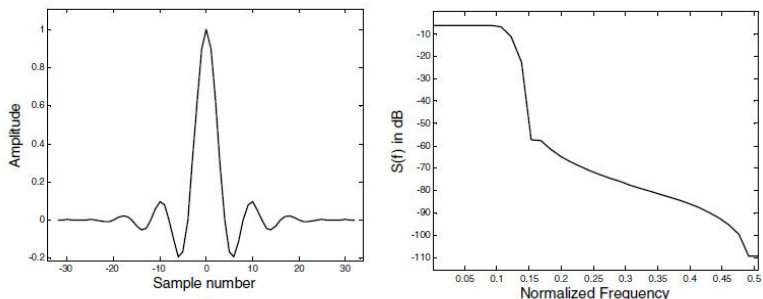

(a) RC pulse (b) PSD of RC pulse

*Wavelet 1*: Mother wavelet
*Wavelet 1.5*: Mother wavelet + 1. Dyadic Expansion
*Wavelet 1.75*: Mother wavelet + 1. & 2. Dyadic Expansion

| System | Spectral Efficiency (Base band) b/S/Hz | Spectral Efficiency (Pass band) b/S/Hz |
|---|---|---|
| RC | 0.83 | Not Applicable |
| Wavelet 1 | 0.7 | 0.85 |
| Wavelet 1.5 | 1.01 | 1.10 |
| Wavelet 1.75 | 1.12 | 1.22 |

# Wavelet Modulation (3/3)
## System 2: Wavelet Shift Keying (WSK)

- Replace the classical oscillator by a waveform generator that produces a basic mother wavelet.

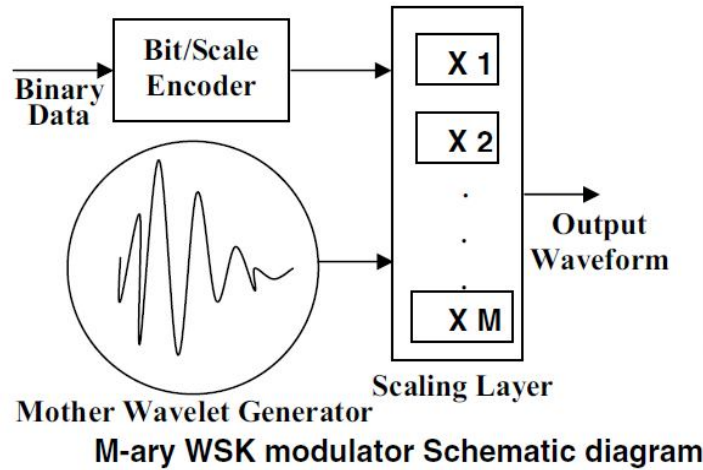

M-ary WSK modulator Schematic diagram

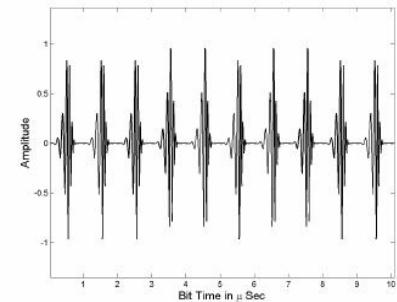

Part of binary BWSK bit stream

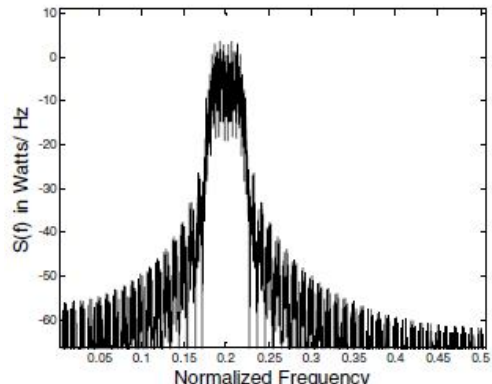

PSD of BPSK with RC filtering

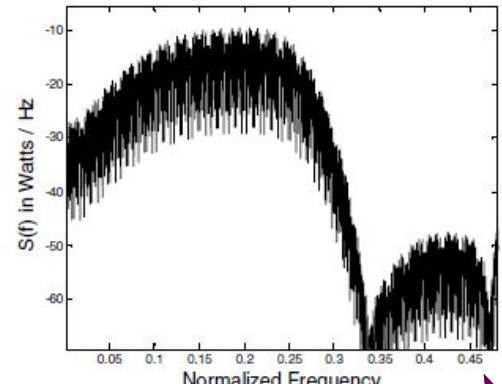

PSD of BWSK signal

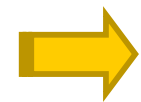

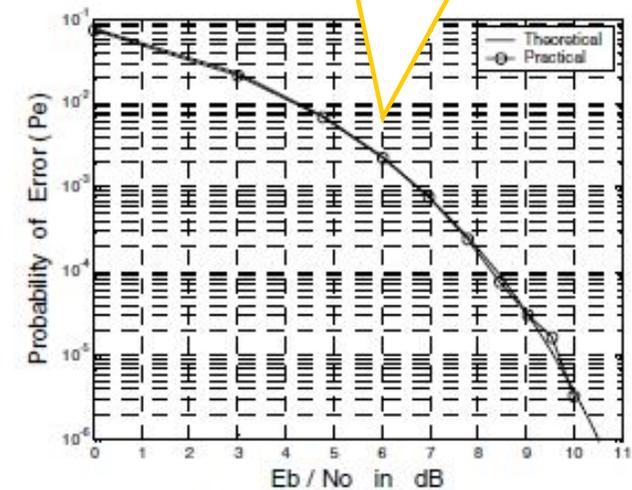

Performance of BWSK

Similar Performances

Transmit other user's information on the same BW.

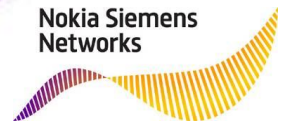

# Bandwidth Efficiency
## Spectra of Different Wavelet Families

- Wavelet diversity can be exploited to find the wavelet function with desired bandwidth characteristics.

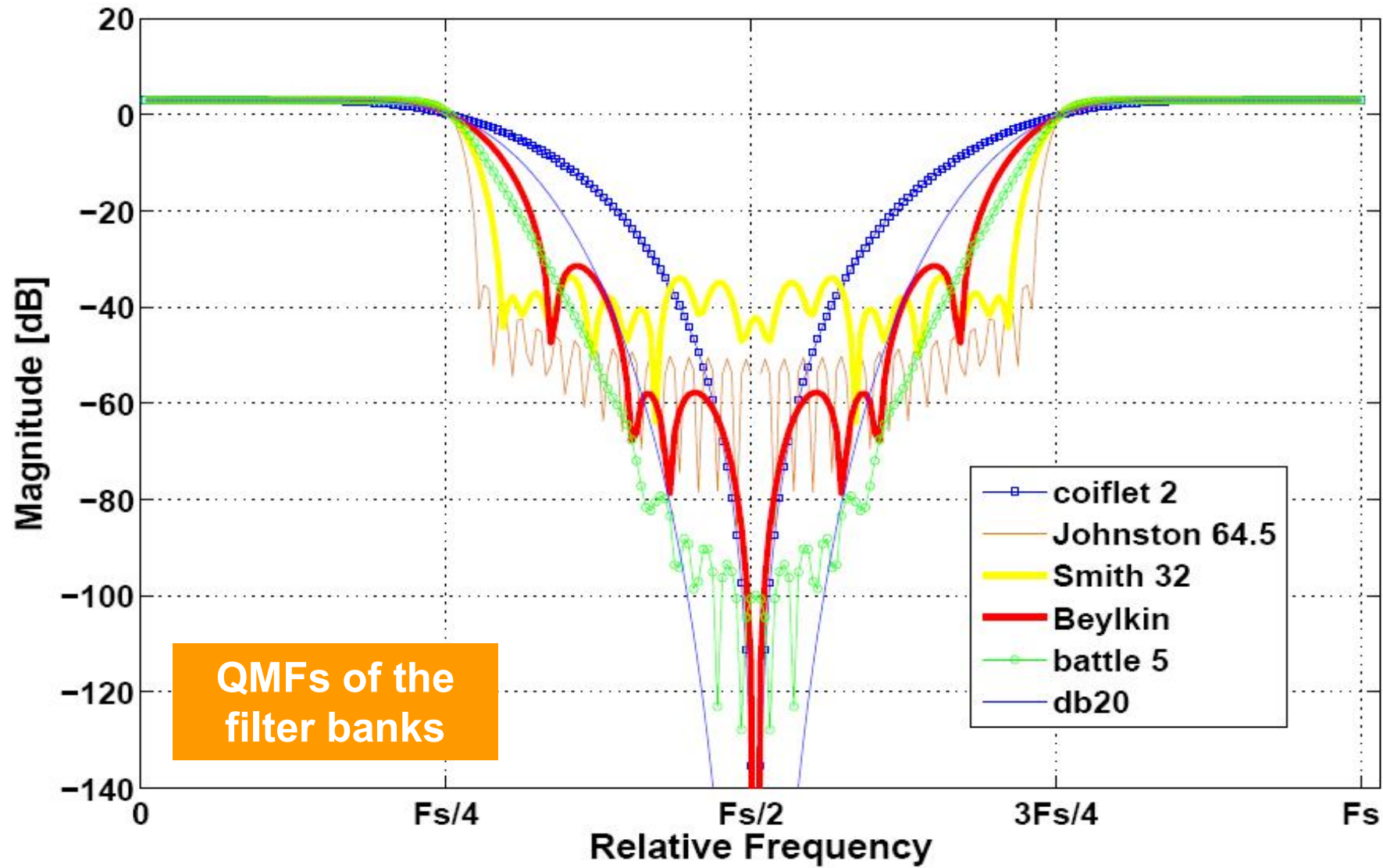

QMFs of the filter banks

# Bandwidth Efficiency
## Impact of Receiver Filter Bandwidth

- Error Vector Magnitude (EVM) is defined as:

$$\text{EVM} = \frac{1}{L_s} \sum_{i=1}^{L_s} \frac{|\underline{e}_i|^2}{|\underline{d}_i|^2}$$

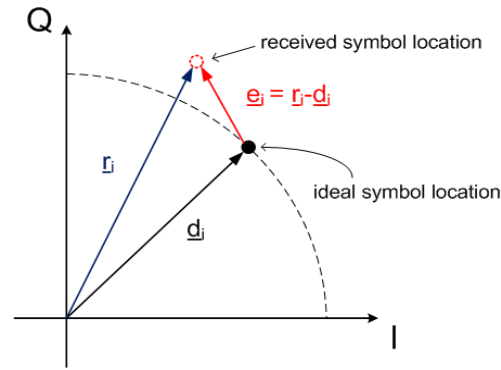

- In order to investigate the impact OFDM symbols (*N*=512 SCs with oversampling factor of 2) are generated using Fourier transform and Wavelet Packet transform (Daubechies - db20).

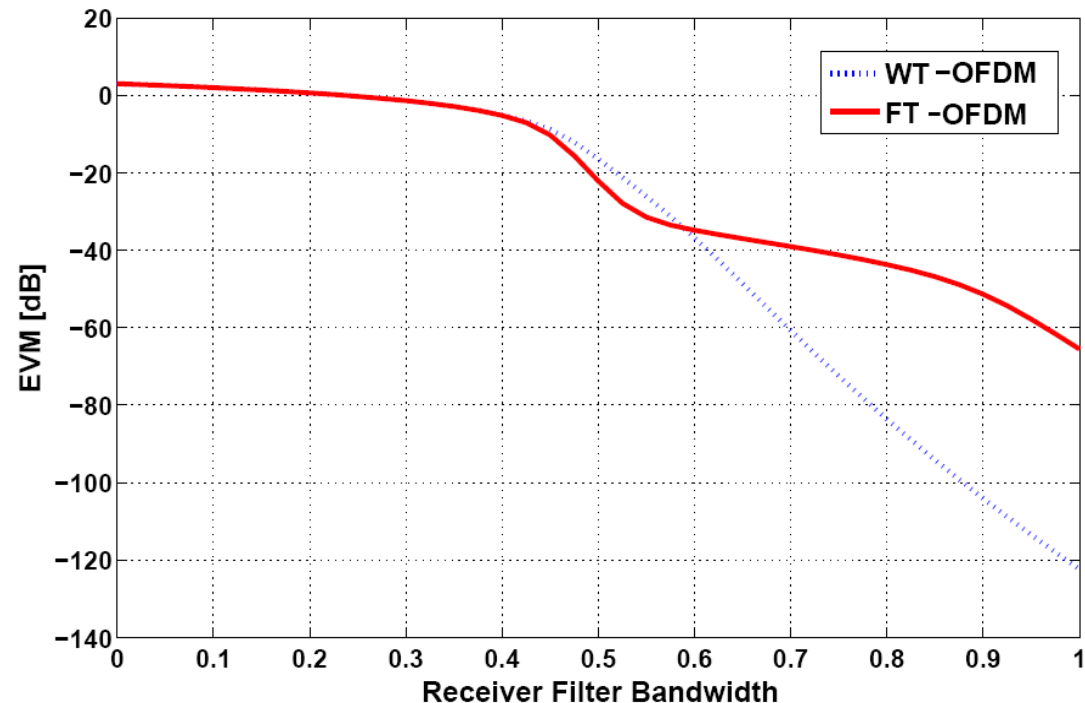

- Introduction

- Wavelet Modulation

- Bandwidth Efficiency
    - Spectra of Different Wavelet Families
    - Impact of Receiver Filter Bandwidth

- **The Modified Gaussian: A Novel Wavelet**

- Wavelets in SC-FDMA

- Conclusion

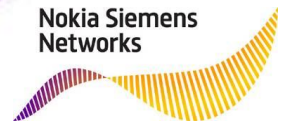

# The Modified Gaussian: A Novel Wavelet

- The modified Gaussian has very low side lobes and its spectral efficiency is controlled by a parameter.
- A so-called *orthogonalization trick* is applied to obtain this wavelet.

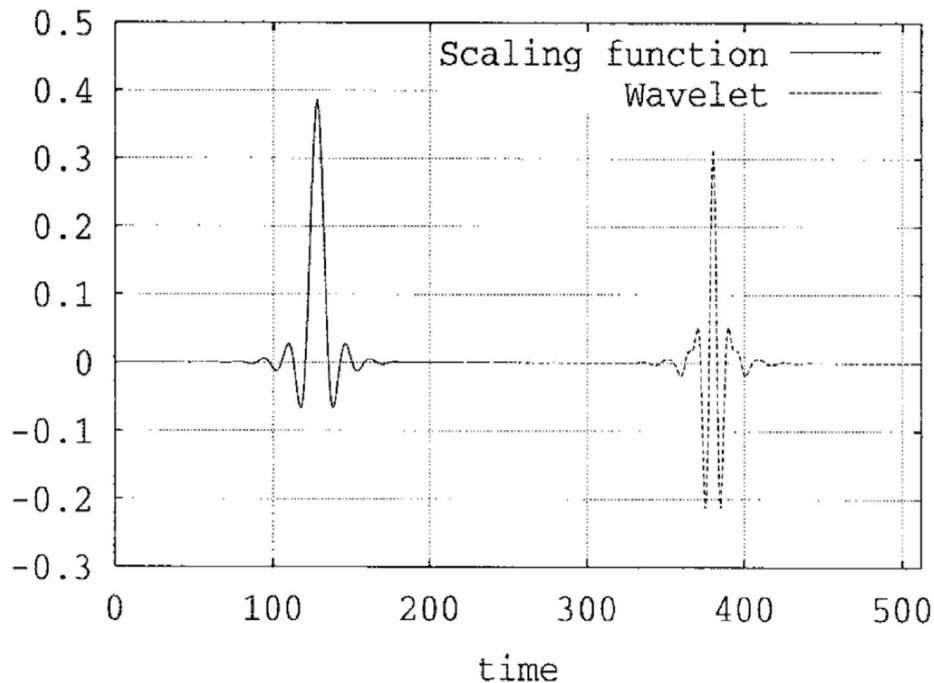
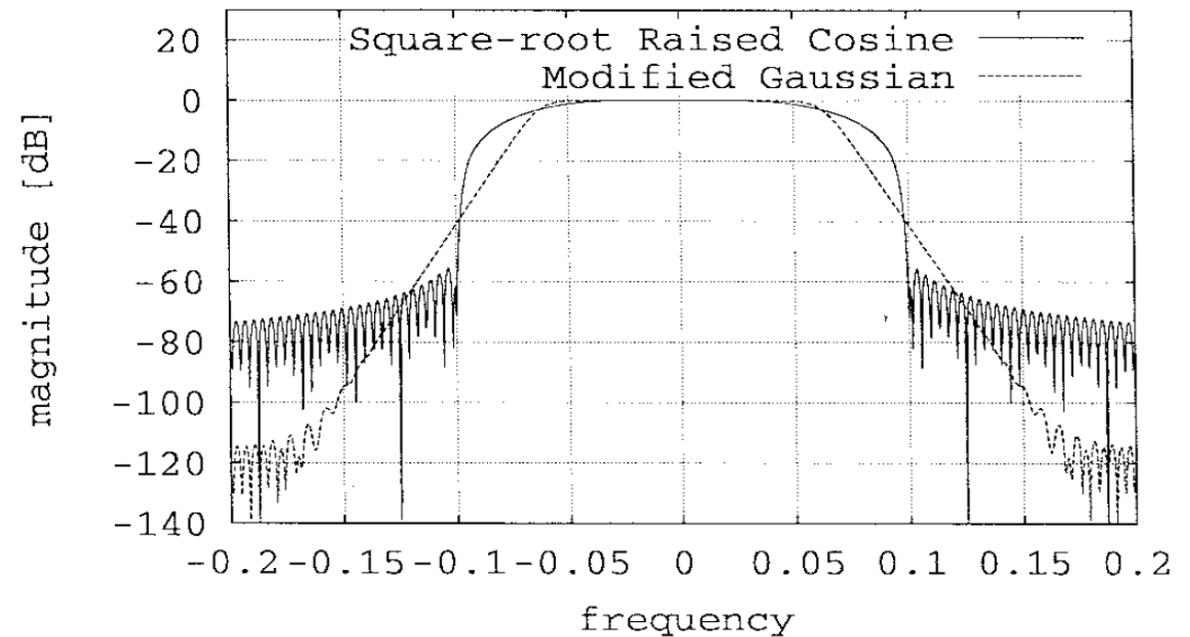

$$\phi(t) \xrightarrow{FT} \Phi(f) = \frac{e^{-\sigma^2 T^2 (2\pi f)^2}}{\sqrt{\sum_{l \in \mathcal{Z}} e^{-8\sigma^2 T^2 \pi^2 (f+l/T)^2}}}$$

- Introduction

- Wavelet Modulation

- Bandwidth Efficiency
    – Spectra of Different Wavelet Families
    – Impact of Receiver Filter Bandwidth

- The Modified Gaussian: A Novel Wavelet

- **Wavelets in SC-FDMA**

- Conclusion

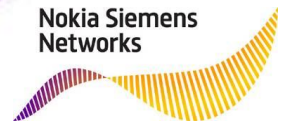

# Wavelets in SC-FDMA (1/3)

- One drawback of the multicarrier modulation system is the high peak-to-average power ratio (PAPR) which in turn decreases the power efficiency.

- The PAPR of an OFDM symbol can be up to $10*\log10(N)$ in the worst case given $N$ as the number of subcarriers.

- Being a multicarrier modulation scheme Wavelet Packet Transform based OFDM also suffers from the high PAPR.

$$CCDF(PAPR_0) = Prob[PAPR > PAPR_0]$$

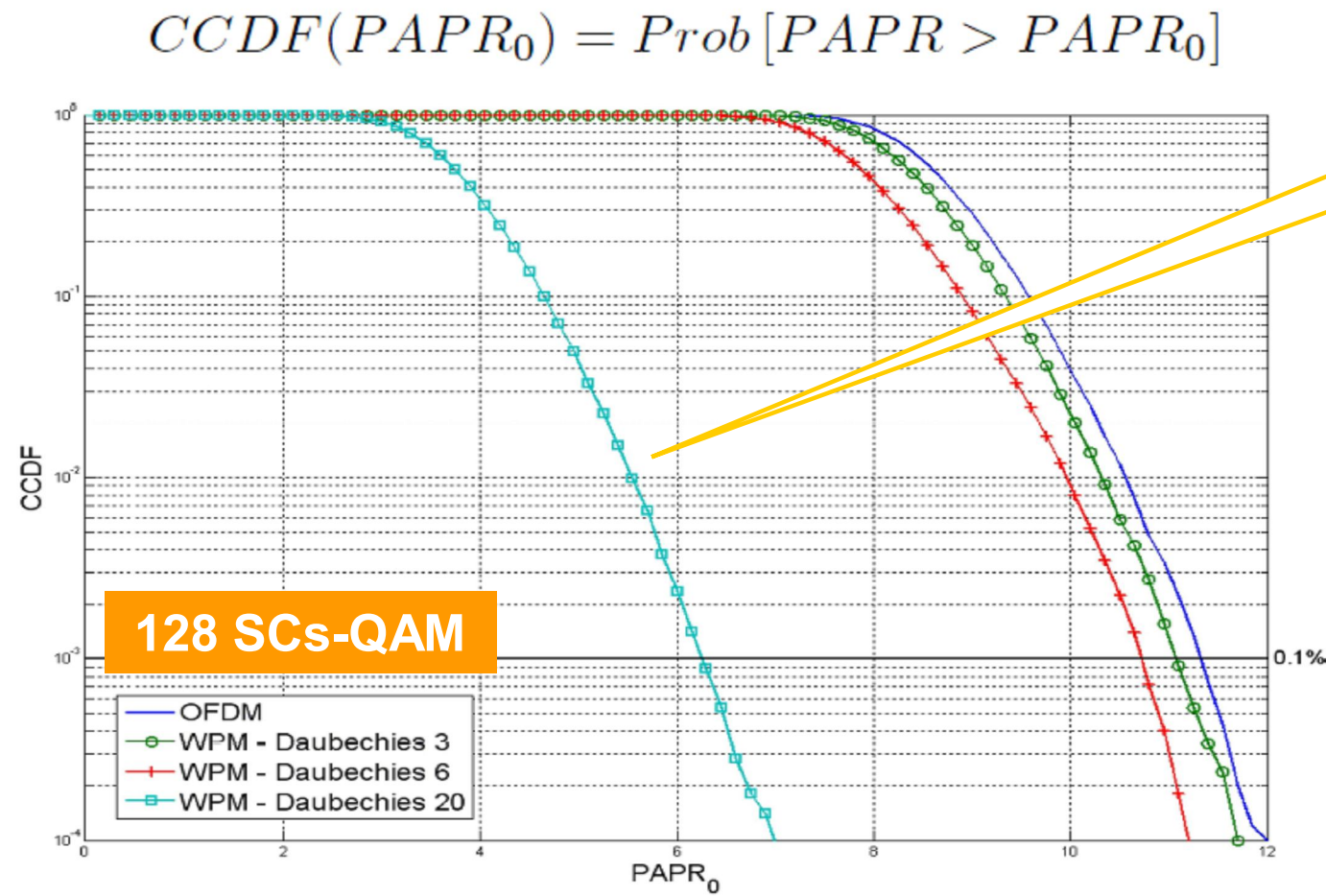

**Decreased PAPR for increased wavelet order**

128 SCs-QAM

# Wavelets in SC-FDMA (2/3)

- In LTE-Advanced Single Carrier (SC)-FDMA is utilized in Uplink to decrease the PAPR and hence to increase the power efficiency.

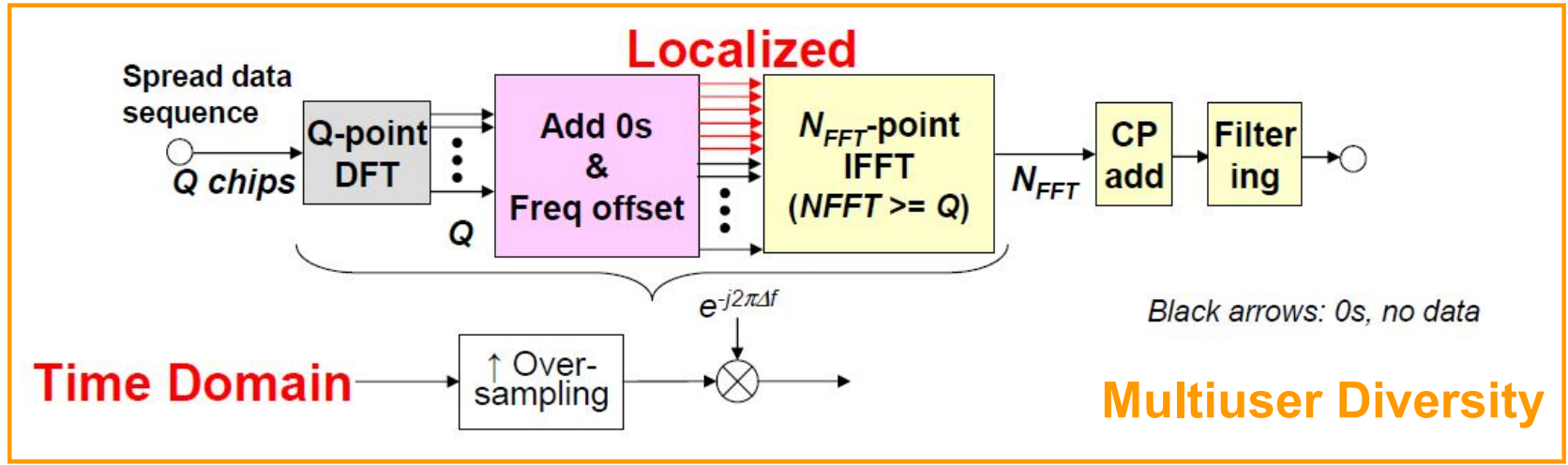

**Multiuser Diversity**

- What about SC-FDMA based on the Wavelet Packet Modulation?  SURE!

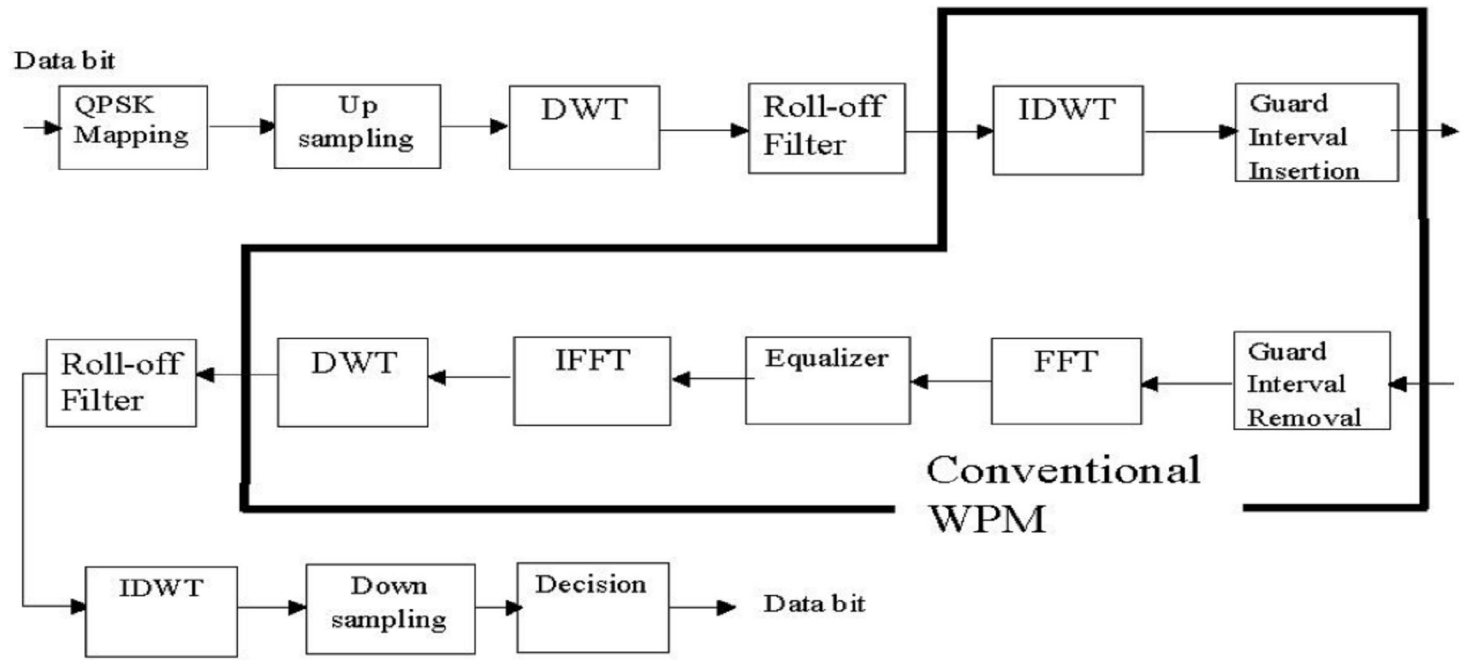

# Wavelets in SC-FDMA (3/3)

- Simulation results reveal better PAPR performance for the wavelet based SC-FDMA compared to conventional Fourier transform based SC-FDMA.

|  | SC-WPM | SC-OFDM | WPM | OFDM |
|---|---|---|---|---|
| Modulation method | QPSK | | | |
| Mother wavelet | Haar | | Haar | |
| Subband level | Level-9 | | Level-6 | |
| Number of FFT points | | 512 | | 64 |
| Roll of factor | 0.0 0.2 0.5 | | | |
| Over sampling ratio | 4 | | | |
| Transmission line | 10-path static fading channel | | | |
| Symbol length | T | | | |
| Guard interval length | T/8 | | | |
| Channel estimation | Perfect | | | |

**10-path Fading Environment**

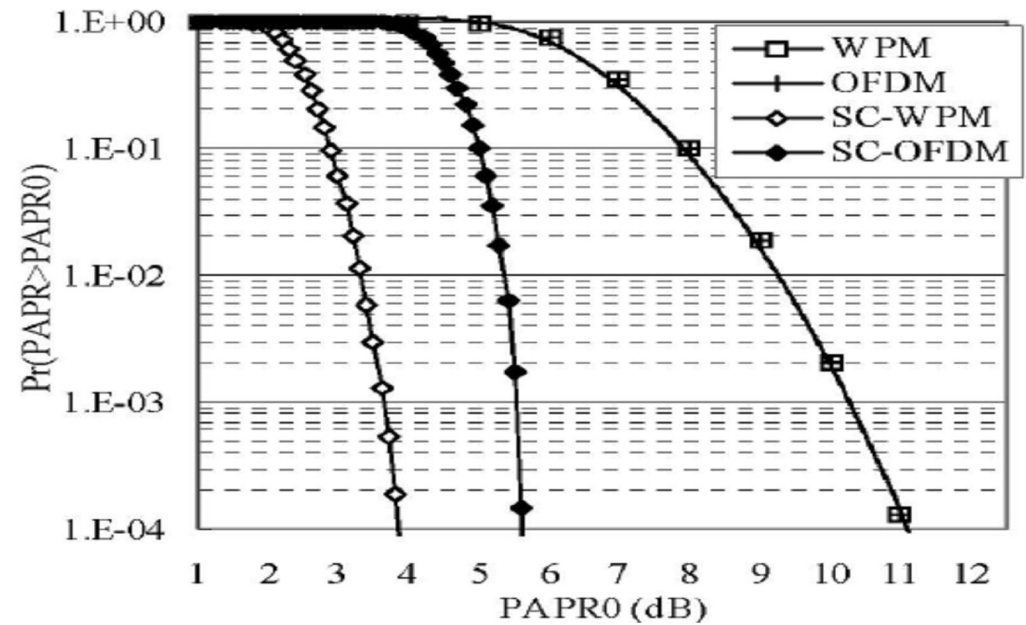

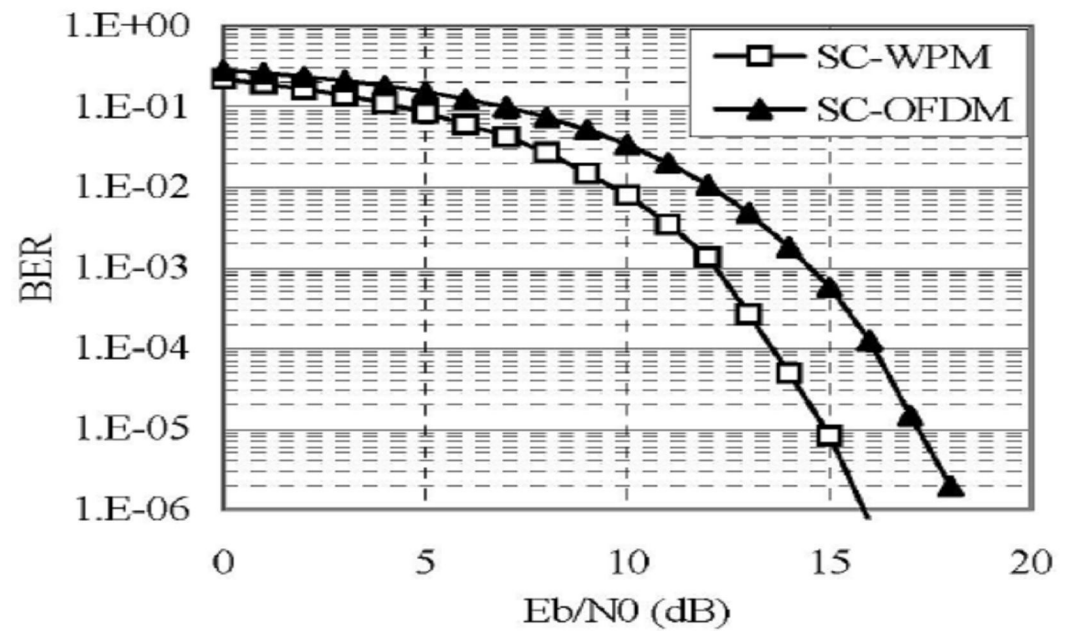

- Introduction

- Wavelet Modulation

- Bandwidth Efficiency
    - Spectra of Different Wavelet Families
    - Impact of Receiver Filter Bandwidth

- The Modified Gaussian: A Novel Wavelet

- Wavelets in SC-FDMA

- Conclusion

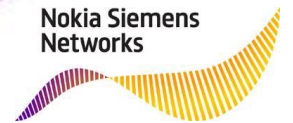

# Conclusion

- **Wavelet diversity and high spectral efficiency properties of wavelets make them a good candidate in single carrier communications.**

- **The modified Gaussian wavelet is promising since it has low side-lobes and its spectral efficiency can be controlled by a parameter.**

- **Similar to conventional Fourier Transform based SC-FDMA, wavelet transform based SC-FDMA promises good PAPR performance.**

# Thank you for your attention!

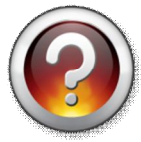 **Q & A**

# BACK-UP

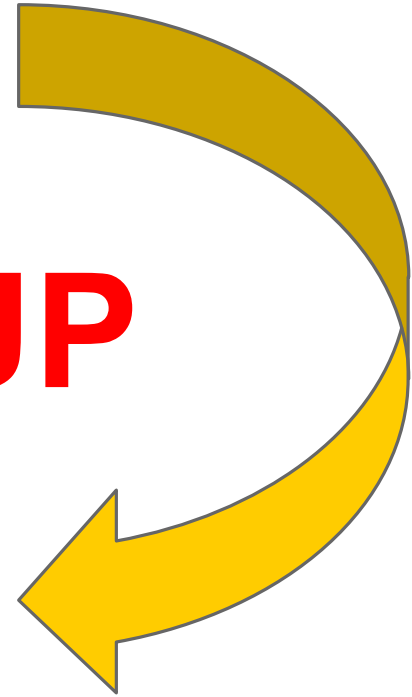

# Wavelet Analysis

**I.  Requirements**

$$C_\psi = \int_0^{+\infty} \frac{|\Psi(w)|^2}{w} dw < +\infty$$ : **Admissibility Condition** where U(w) is the Fourier transform of u(t), the mother wavelet

- **Regularity conditions:** Wavelets should be locally smooth and concentrated in both the time and frequency domains
- **Children Wavelets:** Should form an orthonormal basis of **L²(R)**.

$$u_{ab}(t) = \frac{1}{\sqrt{2}} \cdot u(\frac{t-b}{a})$$  : a=$2^m$ (scaling) and b=n*$2^m$ (shifting) & (m, n) ∈ **Z**²

**II.  Wavelet Transform**

$$W_f(u, s) = \int_{-\infty}^{+\infty} f(t) \frac{1}{\sqrt{s}} \psi^*(\frac{t-u}{s}) dt$$

with $s \in \mathbb{R}^+$ and $s \neq 0$, $u \in \mathbb{R}$

$$f(t) = \frac{1}{C_\psi} \int_{-\infty}^{+\infty} W_f(u, s) \frac{1}{\sqrt{s}} \psi(\frac{t-u}{s}) \frac{ds}{s^2}$$

$$f(t) = \sum_{m \in \mathbb{Z}^2} \sum_{n \in \mathbb{Z}^2} W_f(m, n) \psi_{m,n}(t)$$

Aalto University School of Electrical Engineering — Nokia Siemens Networks

# Wavelet Analysis

III. **DWT (Discrete Wavelet Transform)**

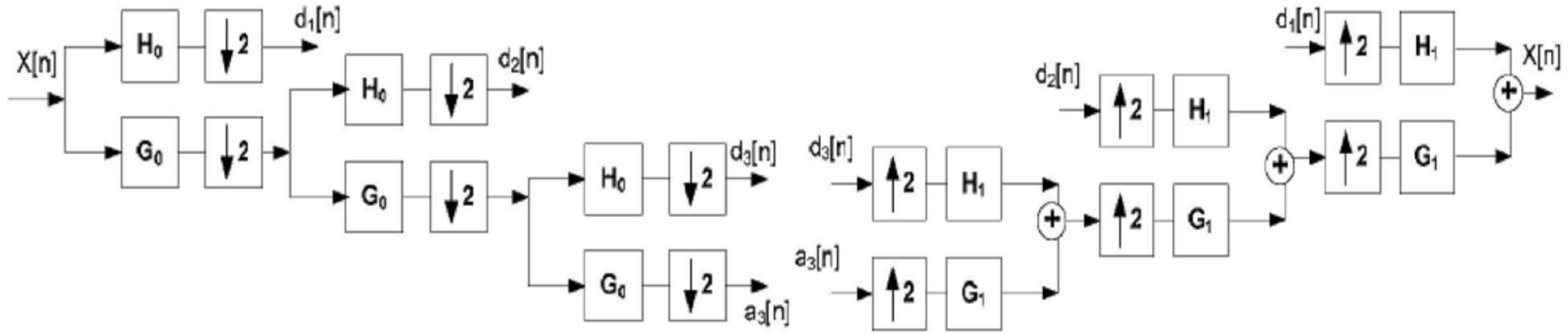

$$G_0(-z)G_1(z) + H_0(-z)H_1(z) = 0$$

(1) **Aliasing-free condition**

$$G_0(z)G_1(z) + H_0(z)H_1(z) = 2z^{-d}$$

(2) **Amplitude Distortion has amplitude of one**

$$h[L-1-n] = (-1)^n \cdot g[n]$$

: **$L$ is the filter length**

# Wavelet Analysis

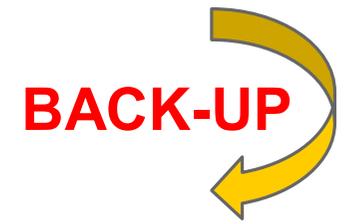

**IV.    DWT Tree Pruning**

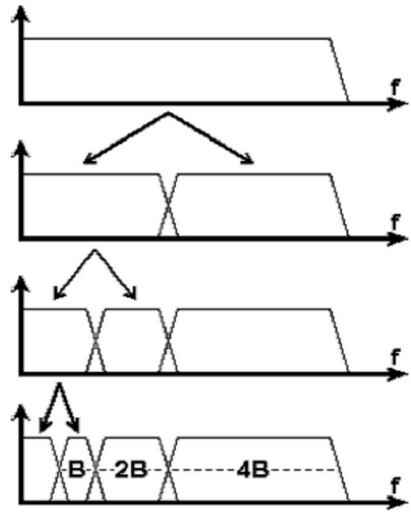
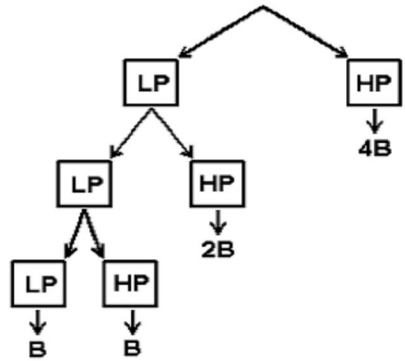
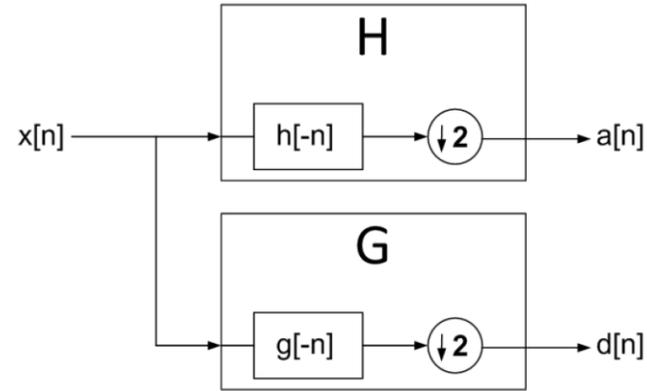

$$a[k] = \sum_{n=-\infty}^{+\infty} h[n-2k]x[n] = x[k] * h[-2k]$$

$$d[k] = \sum_{n=-\infty}^{+\infty} g[n-2k]x[n] = x[k] * g[-2k]$$

**V.    WPT Tree Pruning**

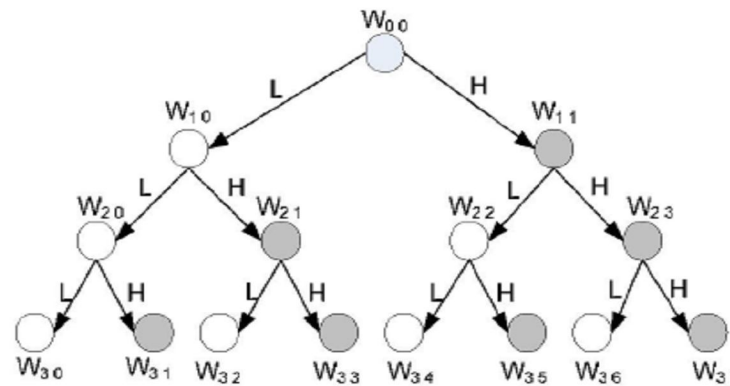

# Wavelet Analysis

**VI.** **FWT Requirements**

$$\frac{1}{\sqrt{2}} \phi(\frac{t}{2}) = \sum_{n=-\infty}^{+\infty} h[n]\phi(t-n)$$

with

$$h[n] = \frac{1}{\sqrt{2}} \left\langle \phi(\frac{t}{2}), \phi(t-n) \right\rangle$$

$$\Phi(w=0) = \int_{-\infty}^{+\infty} \phi(t)dt = 1$$

**Father Wavelet = Scaling Function**

$$\frac{1}{\sqrt{2}} \psi(\frac{t}{2}) = \sum_{n=-\infty}^{+\infty} g[n]\phi(t-n)$$

with

$$g[n] = \frac{1}{\sqrt{2}} \left\langle \psi(\frac{t}{2}), \phi(t-n) \right\rangle$$

$$\Psi(w=0) = \int_{-\infty}^{+\infty} \psi(t)dt = 0 \quad \text{with} \quad \|\psi\| = 1$$

**Mother Wavelet = Wavelet Function**